\documentclass[
,aps
,pra
,amsmath
,eqsecnum
,amssymb
,amsfonts
,twocolumn
,letterpaper
,10pt
,tightenlines
,superscriptaddress
,eqsecnum
,longbibliography
,notitlepage
]{revtex4-2}

\usepackage{graphicx}
\usepackage{dcolumn}
\usepackage{bm}
\usepackage{hyperref}
\hypersetup{colorlinks = true, linkcolor = [rgb]{0.19411,0.51882,0.667058}, urlcolor = [rgb]{0.125490,0.29542,0.1647058}, citecolor = [rgb]{0.75882,0.37411,0.14117}}

\usepackage{amsmath}
\usepackage{mathtools}
\usepackage{setspace}

\graphicspath{{../images/}}
\usepackage[toc,page]{appendix}

\newcommand{\nn}{\nonumber \\}
\usepackage{xcolor}

\usepackage{braket}
\DeclareMathOperator{\tr}{Tr}

\DeclareMathOperator{\Cov}{Cov}

\def\>{\rangle}
\def\<{\langle}

\usepackage[bottom]{footmisc}

\usepackage[normalem]{ulem}
\usepackage{makecell}

\renewcommand{\vec}[1]{\bm{\mathrm{#1}}}
\newcommand{\mat}[1]{\bm{\mathrm{#1}}}
\newcommand{\op}[1]{\hat{#1}}
\newcommand{\opvec}[1]{\hat{\vec{#1}}}

\newcommand{\tp}[0]{\mathrm{T}}

\usepackage{lipsum}

\newcommand{\blk}{\color{black}}

\allowdisplaybreaks

\begin{document}

\title{\texorpdfstring{Limited quantum advantage for stellar interferometry\\via continuous-variable teleportation}{Limited quantum advantage for stellar interferometry via continuous-variable teleportation}}

\author{Zixin Huang$^\natural$}
\thanks{$^\natural$ These authors contributed equally to this work}
\email{zixin.huang@mq.edu.au}
\affiliation{School of Mathematical and Physical Sciences, Macquarie University, NSW 2109, Australia}
\affiliation{Centre for Quantum Software and Information, Faculty of Engineering and IT, University of Technology, Sydney, Australia}

\author{Ben Q. Baragiola$^\natural$}
\email{ben.baragiola@rmit.edu.au}
\affiliation{ Centre for Quantum Computation and Communication Technology, School of Science, RMIT University, Melbourne, VIC 3000, Australia}

\author{Nicolas C. Menicucci}
\affiliation{ Centre for Quantum Computation and Communication Technology, School of Science, RMIT University, Melbourne, VIC 3000, Australia}

\author{Mark M.~Wilde}
\email{wilde@cornell.edu}
\affiliation{School of Electrical and Computer Engineering, Cornell University, Ithaca, New York 14850, USA}

\begin{abstract}
We consider stellar interferometry in the continuous-variable (CV) quantum information formalism and use the quantum Fisher information (QFI) to characterize the performance of three key strategies: direct interferometry (DI), local heterodyne measurement, and a CV teleportation-based strategy. 
In the lossless regime, we show that a squeezing parameter of $r\approx 2$ (18 dB) is required to reach $\approx$ 95\% of the QFI achievable with DI; such a squeezing level is beyond what has been achieved experimentally. In the low-loss regime, the CV teleportation strategy becomes inferior to DI, and the performance gap widens as loss increases. Curiously, in the high-loss regime, a small region of loss exists where the CV teleportation strategy slightly outperforms both DI and local heterodyne, representing a transition in the optimal strategy.
We describe this advantage as limited because it occurs for a small region of loss, and the magnitude of the advantage is also small. 
We argue that practical difficulties further impede achieving any quantum advantage, limiting the merits of a CV teleportation-based strategy for stellar interferometry.

\end{abstract}
\date{\today}
 
\maketitle

\tableofcontents

\section{Introduction}

Interferometry forms the basis for much of astronomical imaging \cite{lawson2000principles,monnier2003optical}. Its performance is limited by diffraction: the resolution is proportional to the aperture of the receiver and inversely proportional to the wavelength -- the ideal instrument is a large-baseline optical interferometer. By combining signals collected across telescope arrays, the achievable resolution is equivalent to that of a large telescope the size of the array's baseline.

In optical interferometric arrays, photons arriving at different telescopes are connected by physical optical links, such as fibres and other optical elements, that bring them together for an interference measurement \cite{monnier2003optical,ten2000technical}. However, optical elements are inherently lossy, and if the telescopes are separated by long distances, bringing the photons together to perform such an interference measurement would result in most of the signal being lost. Bypassing that
requires quantum resources such as entanglement and some form of established coherence between the nodes in the array~\cite{PhysRevLett.109.070503,PhysRevLett.107.270402,PhysRevLett.123.070504,PhysRevLett.129.210502}. 

Several quantum-enhanced protocols \cite{PhysRevLett.109.070503,PhysRevLett.123.070504,PhysRevLett.129.210502,brown2022interferometric} have considered the weak-photon limit, in which photons arriving from the source are shared nonlocally between the telescope sites.
There, pre-distributed and distilled entanglement replaces the lossy optical link, and
discrete-variable measurement protocols are used to estimate the parameters of interest. 
However, these discrete-variable protocols truncate the quantum state at the single or two-photon level, even though states received from astronomical sources are inherently thermal \cite{mandel1995optical}. Therefore, it is natural to consider this problem in the framework of continuous-variable (CV) quantum information \cite{serafini2017quantum,RevModPhys.84.621}, which motivates our work.
 Any imaging task can be translated into a parameter estimation task, for which an essential figure of merit is the quantum Fisher information (QFI) (see, e.g., \cite{SK20} for a review). Here, we take into account transmission loss in the distributed entanglement and quantify the QFI in the presence of this loss.

\begin{figure}
\includegraphics[trim = 0cm 0.0cm 0cm 0cm, clip, width=1\linewidth]{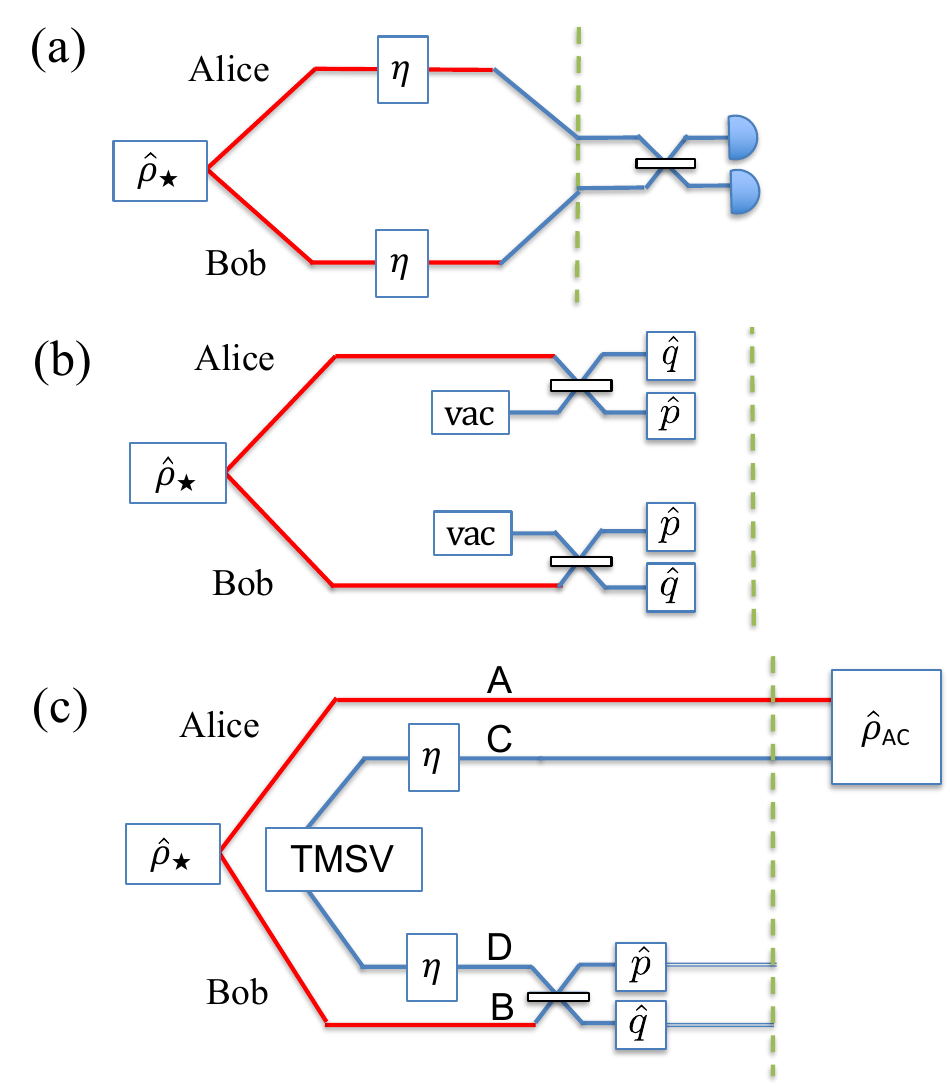} 
 \caption{\label{f:schemes} The three strategies we compare for estimation of parameters in the stellar state $\op{\rho}_\star$.
 (a) direct interferometry (DI), where the two modes of the stellar state are physically brought together for interference. Each mode suffers transmission loss parameterized by $\eta \in [0,1]$. 
 (b) A local strategy, where heterodyne detection is performed separately on the modes held by Alice and Bob, and no loss is incurred.
 (c) A CV teleportation strategy, where a two-mode squeezed vacuum (TMSV) is distributed to Alice and Bob. During distribution, each mode of the TMSV suffers transmission loss parameterized by $\eta$. Bob performs joint homodyne measurements as prescribed by standard CV teleportation and sends his measurement outcomes to Alice.
 }
\end{figure}

Consider a two-site scenario (named Alice and Bob), with each featuring a telescope station such that they are separated by a large distance. As in \cite{PhysRevLett.107.270402}, we model the incoming signal as a correlated thermal state, and the task is to extract the relevant parameters for imaging: the relative phase and the (complex) degree of coherence \cite{mandel1995optical}. To extract these parameters, we need to interfere the modes held by Alice and Bob.
Several schemes can be used, as depicted in Fig.~\ref{f:schemes}: (a) direct interferometry, which requires bringing the signal physically together via an optical link; (b) a spatially ``local" measurement scheme, heterodyne detection -- a phase reference is distributed, but without entanglement; (c)
CV teleportation.

\color{black}
In this work, we scrutinize the performance of these schemes, where we characterize the scheme (c) inspired by CV teleportation \cite{PhysRevLett.80.869}, for which the resource state is a distributed two-mode squeezed vacuum (TMSV).
In the schemes in Refs.~\cite{PhysRevLett.109.070503,PhysRevLett.123.070504,PhysRevLett.129.210502}, the stellar photon is lossless, and the entanglement distribution experiences the loss in place of the stellar photon. 
Here we treat losses in a fair way---i.e.,~either the distributed entanglement or the stellar photon has to travel, and therefore one of them will experience losses. For each scheme we characterize the QFI given a level of loss on the modes. 
Overall, we observe a small quantum advantage afforded by distributed entanglement in the presence of loss; however, achieving it requires measurements that are difficult to realize experimentally, and therefore we deem the quantum gain limited.

The structure of our paper follows. 
In Sec.~\ref{sec:prelim} we briefly review the CV formalism, highlight the key concepts and tools we use from quantum metrology, and describe our model. In Sec.~\ref{sec:results}, we describe our schemes in detail, and we show our results in Sec.~\ref{sec:results1} and \ref{sec:results2}. We conclude in Sec.~\ref{sec:con} with a summary and some directions for future work.

\section{Preliminaries}
\label{sec:prelim}

\subsection{Gaussian formalism}

Consider $n$ bosonic modes described by quadrature operators 
\begin{equation}                     
         \op q_j \coloneqq \tfrac{1}{\sqrt{2}}(\op b_j + \op b_j ^\dagger) \quad \text{and} \quad
         \op p_j \coloneqq \tfrac{-i}{\sqrt{2}}(\op b_j - \op b_j ^\dagger) ,
\end{equation}
whose mode creation and annihilation operators $\op b_j$ and $\op b^\dagger_j$ satisfy $[\op b_j, \op b_k^\dagger] = \delta_{j,k}$. 
The vector of quadrature operators 
$\opvec{x} \coloneqq (\hat q_1,\hat p_1,\ldots, \hat q_n, \hat p_n)^\tp$ 
satisfies
\begin{align}
[\op{x}_j, \op{x}_k] = i \Omega_{jk}, \qquad 
\mat \Omega \coloneqq \openone \otimes \left(
\begin{matrix} 
0 & 1 \\
-1 & 0 
\end{matrix} \right) ,
\end{align}
\noindent where $\openone$ is the $n\times n$ identity matrix. 

Any Gaussian state $\hat \rho$ is entirely specified by its
first and second moments with respect to the quadrature operators \cite{serafini2017quantum,RevModPhys.84.621}---i.e., a mean vector $\vec{r} \in \mathbb{R}^{2n \times 1}$ and covariance matrix $\mat{\sigma}\in \mathbb{R}^{2n \times 2n}$ whose elements are given by
\begin{align}
r_j & \coloneqq \operatorname{Tr}[ \op x_j \op \rho] , \\ 
\sigma_{jk} & \coloneqq \operatorname{Tr} \!\big[ \{ \op x_j-r_j, \op x_k-r_k \} \op \rho \big],
\end{align}
\noindent where $\{\op A ,\op B \} \coloneqq \op A \op B+\op B \op A$ denotes the anticommutator.

\subsection{The stellar state}

\label{sec:model}

The task of imaging can be recast into a parameter estimation problem: by estimating the relevant parameters, we can optimally reconstruct the spatial configuration of the objects of interest. To estimate the spatial configuration of the source, the simplest imaging scenario requires two spatial modes for collecting the signal \cite{Pearce2017optimal}. Optimal parameter estimation using two spatial modes has been shown to surpass the classical diffraction limit of direct imaging, for estimating the separation of sources \cite{PhysRevLett.124.080503}, as well as for detecting secondary sources \cite{zanforlin2022optical}.
We model the incoming stellar signal as a correlated thermal state of light, $\op{\rho}_\star$, that has been multiplexed into frequency bands narrow enough for interferometry. As it is a Gaussian state, $\op{\rho}_\star$ is fully specified by its mean vector and covariance matrix.

We consider a single frequency band whose mean vector and covariance matrix are given by~\cite{mandel1995optical,Pearce2017optimal, PhysRevLett.129.210502}.
\footnote{In those works, the covariance matrix is presented in a basis of creation and annihilation operators. A change of basis to position and momentum gives the form that we use.}
\begin{align} 
\label{eq:star}
\vec r_\star & \coloneqq (
\begin{array}{cccc}
0 & 0 & 0 &0
\end{array})^\tp,
\\
\mat \sigma_\star & \coloneqq 
\left(
\begin{array}{cccc}
 \epsilon +1 & 0 & \gamma  \epsilon  \cos \phi   & -\gamma  \epsilon  \sin \phi  \\
          0  & \epsilon +1                        & \gamma  \epsilon  \sin \phi  &  \gamma  \epsilon  \cos \phi \\
  \gamma  \epsilon  \cos \phi & \gamma  \epsilon  \sin \phi  & \epsilon +1 & 0  \\
 -\gamma  \epsilon  \sin \phi  & \gamma  \epsilon  \cos \phi & 0  & \epsilon +1
\end{array}
\right),
\label{eq:star2}
\end{align}
where we have used the quadrature ordering $(q_A, p_A, q_B, p_B)$, with subscripts referring to Alice~($A$) and Bob ($B$).
The parameter $\phi \in [0 ,2\pi )$ is related to the location of the sources 
and $\gamma \in [0,1]$ is proportional to the Fourier transform of the intensity distribution (shape of the objects) via the van Cittert--Zernike theorem \cite{mandel1995optical}. If $\gamma=1$, the object is a single point source, and $\gamma$ decreases as the size of the object increases.
The parameter $\epsilon \coloneqq  \langle\op{n}_A\rangle + \langle\op{n}_B\rangle$ is equal to the total mean photon number across the two spatial modes. The covariance matrix in
Eq.~\eqref{eq:star2} can be diagonalised with a suitable beam splitter operation, where the eigenvalues are
$ (1 +\epsilon \pm \gamma\epsilon)$; this implies that the mean photon numbers in the two diagonalised modes are 
\mbox{$\frac{1}{2}(\epsilon(1\pm\gamma))$},
and these are both thermal states.

Note that many previous analyses use a single-photon approximation to the state above, which is valid when $\epsilon \ll 1$~\cite{PhysRevLett.109.070503,PhysRevLett.107.270402,PhysRevLett.123.070504}. The Gaussian formalism we employ does not put any restriction on $\epsilon$.

\subsection{Quantum Fisher information}
\label{sec:qfi}

The ultimate precision in parameter estimation 
is specified by the quantum Cram\'er--Rao bound~\cite{caves,caves1} (see also \cite{giovannetti2011advances,giovannetti2006quantum}).
For estimation of a parameter $\theta$ encoded into a quantum state~$ \op \rho_\theta$, the Cram\'er--Rao bound sets a lower bound on the variance $(\Delta \theta)^2 = \langle \theta^2 \rangle - \langle \theta \rangle^2$ of any unbiased estimator~$\theta$.
For unbiased estimators, the quantum Cram\'er--Rao bound establishes that 
\begin{align} \label{eq:var}
 (\Delta  \theta) ^2 \geqslant  \frac{1}{N  J_\theta( \op \rho_\theta)} \, ,
\end{align}
where $N$ is the number of copies of $\hat \rho_\theta$ used and $J_\theta$ is the quantum Fisher information (QFI) associated with the state $\op \rho_\theta$. 
For detailed discussions of the quantum Cram\'er--Rao bound, see Refs.~\cite{holevo2011probabilistic,hayashi2008asymptotic,suzuki2020quantum,yang2019attaining,hayashi2023tight}.

If there are multiple parameters we want to estimate, where $\vec\theta = (\theta_1, \theta_2, \dots)$, we can define a \emph{QFI matrix} $\mat{J}$
that quantifies not only the QFI for each parameter (diagonal components) but also for correlated parameters (off-diagonal components).
The matrix elements are given by
\begin{align}
J_{jk}
\coloneqq  \frac{1}{2}\text{Tr}[\op \rho_{\vec{\theta}}(\op L_j \op L_k + \op L_k \op L_j)],
\end{align}
 where $\op L_j$ is the symmetric logarithmic derivative with respect to $\theta_j$~\cite{paris2009quantum}. 

The inverse of the QFI matrix provides a lower bound on the covariance matrix
 $\big[ \Cov(\vec\theta) \big]_{jk} = \braket{\theta_j \theta_k } - \braket{\theta_j}\braket{\theta_k}$,
 \begin{align}
 \Cov(\vec\theta) \geq \frac{1}{N } \vec J^{-1}.
\end{align}
For a single parameter, the Cram\'er--Rao bound is known to be attainable \cite{barndorff2000fisher}.
For multiple parameters, the bound is not always attainable because the optimal measurement operators that saturate the bound for the individual parameters may not commute. Therefore, the parameters may not be simultaneously measurable.

Ref.~\cite{Monras} derived a closed form for the QFI of a Gaussian state for a single parameter~$\theta$. We need a version that gives the QFI matrix for a vector of parameters~$\vec \theta$. For this, we turn to Ref.~\cite{gao2014bounds}. Both results involves some tricky notation, and so we give our own presentation of the final form and then relate it to the references above so that the reader may verify it if desired.

For this, we need to define the following objects. First,  we define $\vec \varsigma$ as a $4N^2$-dimensional column vector obtained by stacking the $2N$ columns of $\mat \sigma$ on top of each other. Explicitly,
\begin{align}
    \vec \varsigma
\coloneqq
    (
    \sigma_{11},\dotsc,\sigma_{(2N)1},
    \;
    \dotsc
    ,
    \;
    \sigma_{1(2N)},\dotsc,\sigma_{(2N)(2N)}
    )^\tp
    .
\end{align}
Equivalently, since $\mat \sigma$ is symmetric, $\vec \varsigma^\tp$ is a row-vector obtained by concatenating the rows of $\sigma$ in order (stacking them). Next, we need the $(4N^2 \times 4N^2)$ matrix
\begin{align}
    \mat M
\coloneqq
    \mat \sigma
    \otimes
    \mat \sigma
    -
    \mat \Omega
    \otimes
    \mat \Omega
    ,
\end{align}
where $\otimes$ is the standard matrix Kronecker product. Note that $\mat M^\tp = \mat M$. Finally, we define
\begin{align}
    \partial_j
\coloneqq
    \frac{\partial}{\partial \theta_j}
    ,
\end{align}
corresponding to the components of the gradient operator with respect to~$\vec \theta$.
This lets us express the QFI matrix elements as %
\footnote{In Ref.~\cite{Monras}, only a single parameter is considered, so $\vec \theta$ has just a single entry, and $\partial_j \mapsto \partial$ [where $A \mapsto B$ means $A$ (in our notation) equals $B$ (in the reference's notation)]. Also, $\mat \sigma \mapsto \Gamma$, $\mat \Omega \mapsto \omega$, and $\vec r \mapsto d$.
Additionally, the vectorized version of~$\Gamma$ has a different notation $\vec \varsigma \mapsto \lvert \Gamma )$, defined by row concatenation (irrelevant for a symmetric matrix; see text), and $\vec \varsigma^\tp \mapsto ( \Gamma \rvert$.
To obtain the multi-parameter result, we consulted Ref.~\cite{gao2014bounds}, wherein the basis for expressing the covariance matrix~$\Sigma$ (their notation) is $(\op a, \op a^\dag)$ for both rows and columns, leaving $\Sigma$ complex symmetric (as opposed to Hermitian). Also, the definition of the entries in a covariance matrix differs from ours by a factor of two. Explicitly, $\mat \sigma \mapsto 2 H \Sigma H^\tp$, where $H = \bigoplus_{k=1}^N 2^{-1/2} \left(\begin{smallmatrix}1 & 1 \\ -i & i\end{smallmatrix}\right)$. With this choice of basis operators, $\mat \Omega \mapsto i H \Omega H^\tp = \Omega$.
These differences mean that $\mat M \mapsto 4 (H \otimes H) \mathfrak M (H^\tp \otimes H^\tp)$, with $\mathfrak M = \Sigma \otimes \Sigma + \Omega \otimes \Omega/4$. The plus sign in~$\mathfrak M$ (instead of minus sign in~$\mat M$) arises from conjugating the final term with $H$. Finally, $\vec r \mapsto H \lambda$, and $\vec \varsigma \mapsto 2(H \otimes H) \Sigma_{\text{(vec)}}$, where~$\Sigma_{\text{(vec)}}$ is the (row- or column-)vectorized version of~$\Sigma$.
}
\begin{align}
\label{qfi_matrix} 
    J_{jk}
&=
    \frac 1 2
    (\partial_j \vec \varsigma)^\tp
    \mat M^{-1}
    (\partial_k \vec \varsigma)
    +
    2
    (\partial_j \vec r)^\tp
    \mat \sigma^{-1}
    (\partial_k \vec r)
    ,
\end{align}
where we have employed the findings of \cite{Monras,gao2014bounds}.
Notice that the symmetry of $\mat M$ and~$\mat \sigma$ allow the labels~$(j,k)$ to be freely exchanged on each expression on the right, ensuring that $\mat J$ is symmetric, as required.

\blk

\section{Estimating stellar parameters}
\label{sec:results}

The problem at hand is the estimation of the two unknown stellar parameters $\phi$ and $\gamma$ in the stellar state specified by Eq.~\eqref{eq:star}. Optimally estimating $\phi $ and $\gamma$ provides complete information on what we can obtain about the source distribution by using two spatial modes.
The QFI calculated directly from the stellar state sets the ultimate limit on the precision of estimators for $\phi$ and $\gamma$ via the Cram\'er--Rao bound.  
Saturating this bound requires finding an optimal positive operator-valued measure (POVM) that achieves the QFI, which is not necessarily a simple task,
even for Gaussian states.

The QFI matrix elements for the incoming stellar state can be found using Eq.~\eqref{qfi_matrix}:
\begin{subequations}
\begin{align} 
\label{eq:ideal_qfi_matrix}
J_{\phi } 
&=\frac{2 \gamma ^2 \epsilon }{2+ \epsilon(1-\gamma ^2) } ,
\\
J_{\gamma } &= \frac{2 \epsilon  \left(2 + \epsilon + \epsilon \gamma ^2 \right)}{\left(1-\gamma ^2\right) \left(4+ 4\epsilon + \epsilon ^2 \left(1-\gamma ^2\right) \right)} ,
\\
J_{\phi\gamma} &= 0,
\end{align}
\end{subequations}
where we label the diagonal elements simply as $J_j$ for convenience.
Even though the QFI matrix is diagonal (i.e., the parameters are independent), we cannot estimate them optimally simultaneously because the symmetric logarithmic derivates for $\phi$ and $\gamma$ do not commute~\cite{Pearce2017optimal, suzuki2020quantum}.

\color{black}
For the problem we consider here, it has been shown that a detection scheme called \emph{direct interferometry} indeed realizes the optimal POVM for both parameters~\cite{Pearce2017optimal}.
In direct interferometry, Alice and Bob's signals are mixed on a 50:50 beamsplitter and then measured with photon-number-resolving detectors.

The question then is: can one perform direct interferometry in practice? If not, how achievable is the Cram\'er--Rao bound in realistic settings? The major practical concern arises from the fact that the stations where Alice and Bob collect their portion of the stellar light are necessarily space-like separated.\footnote{Of less practical concern are limitations on realistic photon-number resolving detectors. However, we do not address that here.}
To implement direct interferometry, which requires a nonlocal measurement (with respect to Alice and Bob), the signals must be brought together. Doing so introduces transmission losses on each arm that directly degrade the stellar state. Loss with transmission parameter $\eta$ amounts to replacing $\epsilon$ by $\eta \epsilon$ in the covariance matrix for the stellar state specified by  Eq.~\eqref{eq:star},
which reduces the QFI compared to the lossless case. 
Lossy DI is depicted in Fig.~\ref{f:schemes}(a).

An alternative strategy is to use local measurements and classical communication \cite{zhou2020saturating}.
Although such strategies have been proven to be inferior (in general) to nonlocal strategies for weak-field interferometry ($\epsilon \ll 1$)~\cite{PhysRevLett.107.270402}, they may still be more practical, they can perform better than DI when losses are included, and they are useful for comparison. We consider a local strategy in which Alice and Bob each perform heterodyne detection individually using a shared phase reference. The practical benefit is that no transmission is required and no loss is incurred---measurements are performed directly on the light collected at the two stations.
 It is known that the classical Fisher information for this local heterodyne strategy is guaranteed to perform suboptimally compared to lossless~DI, due to injection of vacuum noise.  
 Nevertheless, this local strategy will outperform DI in the high-loss limit as the baseline becomes arbitrarily large: in that limit, the signal is almost completely lost for DI,  whereas the local measurement is effectively lossless. The local heterodyne strategy is depicted in Fig.~\ref{f:schemes}(b).

In this work, we consider a third strategy based on CV quantum teleportation~\cite{PhysRevLett.80.869}, as mentioned in \cite{PhysRevLett.109.070503} and also considered in more depth recently in \cite{wang2023astronomical} (see also \cite{pirandola2015advances,furusawa1998unconditional} for more background on CV teleportation). In this scheme, Alice and Bob each possess one mode of a TMSV state distributed to them from some central station in addition to their respective portions of the stellar state. Bob mixes his two local modes on a beamsplitter and measures the position quadrature of one mode and the momentum quadrature of the other. He sends his classical measurement outcome to Alice, who uses it to undo an outcome-dependent displacement. This completes the teleportation of Bob's share of the stellar state onto Alice's share of the TMSV. At that point, Alice has access to the full stellar state and can measure it locally in any way she likes---nonlocal POVMs are not required. The teleportation strategy is depicted in Fig.~\ref{f:schemes}(c).

Noise enters this protocol in two ways. First, finite squeezing \cite[Chapter~21]{mandel1995optical} in the TMSV coherently degrades the teleported state. Current state-of-the-art single-mode and two-mode squeezers achieve 15~dB~\cite{PhysRevLett.117.110801} and 10~dB squeezing~\cite{eberle2013stable}, respectively. 
Even as technology improves, some level of noise due to finite squeezing is inevitable due to energy constraints. Second, regardless of the squeezing level, distributing the TMSV to Alice and Bob incurs transmission loss that must be accounted for. This loss arises for the same reason that it occurs for DI---Alice and Bob are distant from each other. 
We note that there may be methods to improve the quality of the shared entangled state. However, these are beyond the scope of our analysis, as entanglement distillation procedures rely on non-Gaussian measurements and post-selection~\cite{PhysRevLett.102.120501,PhysRevA.84.062309,PhysRevA.91.063832,takahashi2010entanglement,PhysRevLett.98.030502}.

\subsection{Teleportation strategy} \label{sec:teleportationstrategy}

The teleportation strategy makes use of a lossy TMSV in modes $C$ and $D$, for which the mean vector and covariance matrix are
\begin{align}
\vec r_\text{TMSV} & \coloneqq 
\left(
\begin{array}{cccc}
0 & 0 & 0& 0
\end{array}\right)^\tp, \nn
\mat \sigma_\text{TMSV} & \coloneqq 
\left(
\begin{array}{cc}
 c \openone_2 & s \mat \sigma_z  \\
 s \mat \sigma_z & c \openone_2  \\
\end{array}
\right),
\end{align}
where $\mat{\sigma}_z$ is the Pauli-$Z$ matrix, and
    \begin{subequations} \label{csparameters}
        \begin{align}
        c &\coloneqq  \eta \cosh(2r) + (1-\eta) , \\
        s &\coloneqq \eta\sinh(2r) ,
        \end{align}
    \end{subequations}
include the squeezing and loss through the parameters $r$ and $\eta$, respectively.
The squeezing in a TMSV 
is often characterized by the measured variance in the squeezed two-mode quadratures $\Delta^2 = \frac{1}{2}e^{-2r}$ (squeezed for $r>0)$ often reported in decibels (dB): $(\Delta^2)_\text{dB} = -10 \log_{10} 2 \Delta^2 = 20 r \log_{10} e$. 
\color{black}
Since the distributed TMSV has finite squeezing and experiences loss, the teleportation will not be perfect. 

The teleportation protocol proceeds by Bob mixing modes $D$ and $B$ on a beamsplitter and then measuring them in orthogonal quadrature bases, obtaining outcomes $\vec{m} \coloneqq (m_q, m_p)^\tp$ with probability density~\cite{serafini2017quantum} 
\begin{align}
\label{eq:r_m}
p(\vec{m}) &\coloneqq\frac{\exp\left(- \frac{\vec{m}^\tp \vec{m}}{(1+c+\epsilon)} \right)}{\pi (1+c+\epsilon)}
.
\end{align}
Note that $p(\vec{m})$ does not depend on the parameters $\phi$ and $\gamma$, because they do not appear in the reduced state at modes $B$ and $D$---all that Bob has access to.
\noindent

After the quadrature measurements on modes $B$ and~$D$, modes $A$ and $C$ (held by Alice) are projected onto a conditional Gaussian state $\op{\rho}_{\phi,\gamma}^{\vec{m}}$, where the subscripts emphasize dependence on the parameters, and the superscript labels the measurement outcomes $\vec{m}$. The state has
mean vector $\vec{r}_{AC} = (\vec{r}_A , \vec{r}_C)^\tp$, where
\begin{align}
\begin{split} \label{eq:r_a}
\vec{r}_A & \coloneqq
-\frac{1}{s} 
\left(
\begin{array}{cc}
 \mu & \nu \\
 \nu & -\mu
\end{array}
\right) \vec{m} , \\ 
\vec{r}_C & \coloneqq 
s (1+\epsilon +c) \mat \sigma_z
\vec{m},
\end{split} 
\end{align}
and covariance matrix
\begin{equation}  \label{eq:covac}
 \mat \sigma_{AC} \coloneqq  
%
\left(
\begin{array}{cccc}
 1+\epsilon -  \kappa & 0    & \mu & -\nu \\
 0           & 1+\epsilon - \kappa   & \nu & \mu \\
  \mu        & \nu &  c- \lambda & 0 \\
 -\nu        &  \mu & 0 & c-\lambda  \\
\end{array}
\right), 
%
\end{equation}
where
\begin{align}
    \kappa & \coloneqq \frac{\gamma^2 \epsilon^2}{1+c+\epsilon}, \qquad \lambda \coloneqq \frac{s^2}{1+c+\epsilon}, \\
    \mu & \coloneqq \frac{\gamma  s \epsilon  \cos \phi }{1+c+\epsilon}, \qquad
 \nu \coloneqq \frac{\gamma  s \epsilon  \sin \phi }{1+c+\epsilon }.
\end{align}
Details of this calculation are given in Appendix~\ref{app:A}.
An important observation is that both the covariance matrix~$\mat \sigma_{AC} $ and the mean vector $\vec{r}_{AC}$ of the post-measurement state carry information about the unknown parameters. However, as is always the case for Gaussian conditioning, only the mean depends on the measurement outcomes~\cite{serafini2017quantum}.
Often for Gaussian states, the mean plays no role in quantities of interest; however, in our setting, the mean explicitly appears in the QFI via the fidelity in Eq.~\eqref{eq:fid1} and cannot be ignored. 

At this point, the teleportation protocol is not complete because Alice would conventionally wait for Bob to send his measurement results and then perform an appropriate displacement of her state. 
Since the QFI is invariant under  unitary transformations that do not depend on the parameters, this step need not be explicitly performed for our comparisons.  
However, it is important to stress that the optimal POVM will, in general, depend on Bob's measurement outcomes, as is evident in Eq.~\eqref{eq:covac}: the mean of the post-measurement state held by Alice explicitly depends on Bob's classical outcomes---without them, she has no information about it; see Eq.~\eqref{eq:r_a}. 
The purpose of teleporting the full stellar state to Alice is so that she can apply the POVM that extracts parameter information local to her. 
Holding her state while waiting for Bob's outcomes requires a quantum memory, which can lead to additional noise. For example, if Alice uses a fiber delay, the memory losses will be comparable to transmission losses over the full distance between Bob and Alice, $\sim\eta^2$, which provides no advantage. To perform better, Alice could use another type of high-efficiency quantum memory; such analysis is beyond the scope of this work.

In a single shot, the post-measurement state depends on the measurement outcome $\vec{m}$, which occurs with probability density $p(\vec{m})$. 

The Cramer--Rao bound dictated by the QFI in Eq.~\eqref{eq:var} is achievable in the limit of large $N$ \cite{lehmann2006theory}. This means that many copies of the state will be required to perform the estimation properly, and a proper accounting for this requires an ensemble average over Bob's outcomes.
We write this ensemble as follows:
\begin{align}\label{eq:qfi_ensemble}
\hat{\Phi}_{\phi,\gamma} \coloneqq \{(
p(\vec{m}),
 \op \rho_{\phi,\gamma}^{\vec{m}})\}_{\vec{m}},
\end{align}
and apply Eq.~(5.45) from Ref.~\cite{Katariya2021}, which gives the elements of the QFI matrix for the ensemble, 
\begin{align} \label{eq:ensembleQFI}
J_{jk}( \op{\Phi}_{\phi,\gamma}) & = 
J_{jk} \big( p(\vec{m}) \big) 
+       \int d\vec{m}~ 
                          p(\vec m)
                          J_{jk}( \op \rho^{\vec{m}}_{\phi,\gamma})
                          \nn
        &=\int d\vec{m}~ 
                          p(\vec m)
                          J_{jk}( \op \rho^{\vec{m}}_{\phi,\gamma}).
\end{align}
The first term on the right-hand side of the first line is the classical Fisher information for parameter $\phi$ or $\gamma$ or correlations between the two \color{black} embedded in the probability distribution $p(\vec{m})$, and the second term on the first line is the quantum contribution: the QFI for each conditional state $J_{jk}( \op \rho^{\vec{m}}_{\phi,\gamma})$
weighted by $p(\vec m)$. In our case, the probability density in Eq.~\eqref{eq:r_m} 
does not depend on $\phi$ or $\gamma$, and so the first term does not contribute to the QFI for either parameter, thus leading to the second line above.

We use Eq.~\eqref{qfi_matrix} to find the QFI matrix elements for the conditional state corresponding to outcome $\vec{m}$, and then we average over the outcomes using Eq.~\eqref{eq:ensembleQFI}. This gives the average QFI matrix for the teleported state:
\begin{align} \label{eq:qfi_tel}
\mat J^{ \text{tel} }( \op{\Phi}_{\phi,\gamma}) = 
\left(
\begin{array}{cc}
 J^\text{tel}_{\phi} & 0 \\
0 & J^\text{tel}_{\gamma} \\
\end{array}
\right)
.
\end{align}
The matrix is diagonal ($J^\text{tel}_{\phi \gamma} = 0$), and we refer the reader to Appendix~\ref{sec:qfi_matrix} for analytical expressions for $J^\text{tel}_{\phi}$ and $J^\text{tel}_{\gamma}$, which are long and unwieldy.

\color{black}
We are now ready to compare the QFI for $\gamma $ and $\phi$ of the different schemes.

\subsection{Comparing the strategies}

The three strategies---direct interferometry (DI), local heterodyne, and teleportation-based---are depicted in Fig.~\ref{f:schemes}. We compare them by calculating the QFI in Eq.~\eqref{eq:J}  for $\phi$ and~$\gamma$, which sets the limit on estimating these parameters through the Cramer--Rao bound. (For the local heterodyne strategy, the measurements are fully specified, and so we use the classical FI for this case). A summary of the strategies and parameters we use is as follows:

\begin{itemize}
\item Direct interferometry (DI). The portions of the stellar state collected by  Alice and Bob are brought together before detection. In doing so, each arm experiences loss with transmission parameter $\eta$. The result is a reduction in the QFI---before loss, the QFI is proportional to $\epsilon$ \cite{PhysRevLett.107.270402}, and after loss, it is proportional to $\eta \epsilon$. 
\item Local heterodyne. Alice and Bob perform heterodyne detection separately with a shared phase reference~\cite{PhysRevLett.107.270402}. No loss is incurred on the stellar state before detection.
In the weak-field limit ($\epsilon \ll 1$), the QFI scales as $\epsilon^2$~\cite{PhysRevLett.107.270402}. For stronger fields, it can scale more favourably as $\epsilon$, in which case local heterodyne performs as well as DI. 
\item CV teleportation strategy. A TMSV with squeezing parameter $r$ is prepared and its modes are distributed to Alice and Bob, each incurring loss with transmission parameter $\eta$. The stellar state is lossless. 
Bob mixes his share of the stellar state and TMSV state on a balanced beam splitter and performs quadrature measurements. We calculate the QFI at this point using the tools presented in Sec.~\ref{sec:teleportationstrategy}.
\end{itemize}
\begin{figure}
\includegraphics[trim = 0cm 0.0cm 0cm 0cm, clip, width=1.0\linewidth]{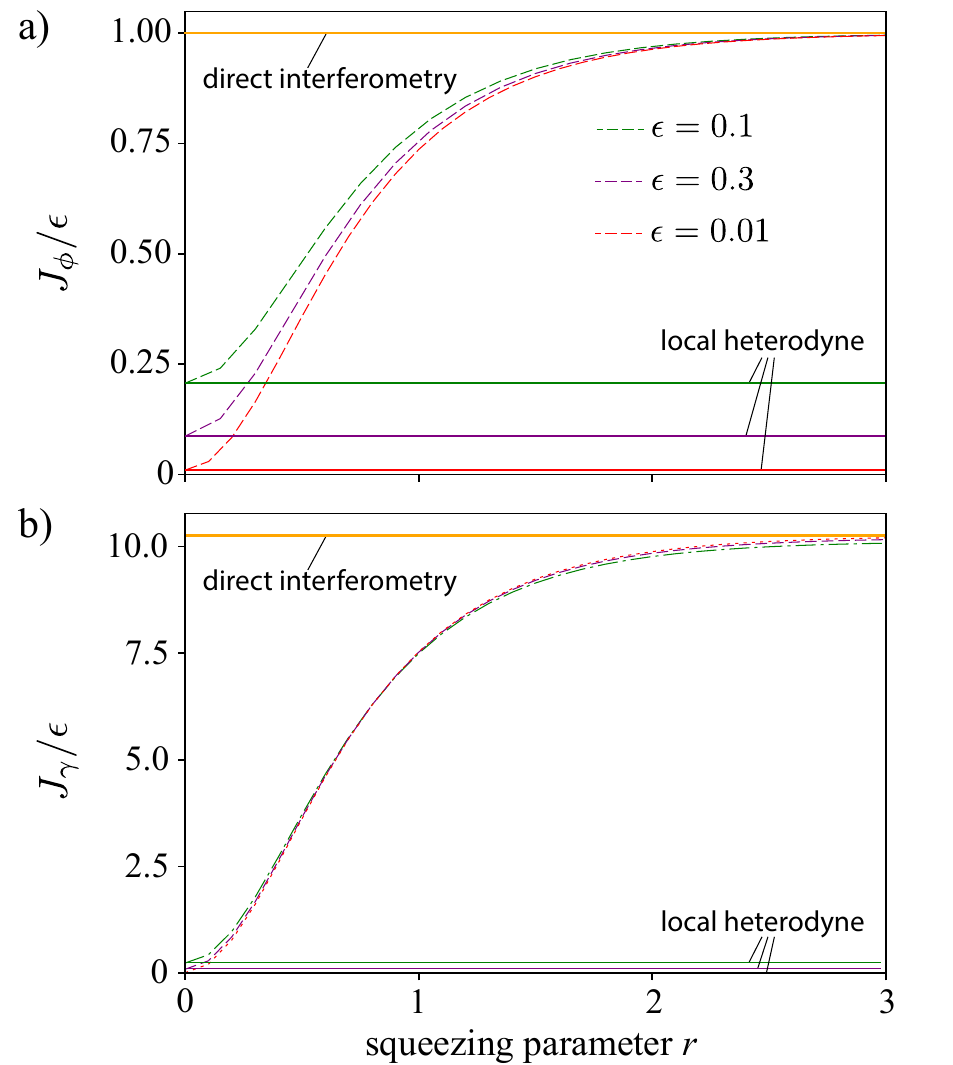} 
\caption{\label{f:qfi_phi_as_function_of_r}
 QFI per photon, $J^\text{tel}_\phi/\epsilon$ and $J^\text{tel}_\gamma/\epsilon$ (dashed lines), in the lossless case ($\eta = 1$) as a function of the squeezing parameter of the TMSV used in the teleportation strategy. The solid yellow lines at the top of each plot are the QFIs of the stellar state $\op{\rho}_\star$, achievable by DI in the lossless case, and the solid horizontal lines at the bottom of each plot are the classical FIs for the local heterodyne strategy for each value of $\epsilon$. (a) QFI $J_\phi$ per photon  when $\gamma = 1$.
 (b) QFI $J_\gamma$ per photon  when $\gamma = 0.95$; note that the QFI $J_\gamma$ does not depend on $\phi$. 
 }
\end{figure}

\subsubsection{QFI for zero loss}
\label{sec:results1}
First, we quantify the performance of the three strategies in the lossless case ($\eta = 1$). The QFI of the teleportation strategy depends on the amount of squeezing in the TMSV, as quantified by the squeezing parameter~$r$. Figure~\ref{f:qfi_phi_as_function_of_r} shows the QFI \textit{per photon}, 
$J_\phi/\epsilon$ and $J_\gamma/\epsilon$, given several values of $\epsilon$. Recall that $\epsilon$ characterizes the mean photon number in the stellar state; see Sec.~\ref{sec:model}. 
For increasing $r$, the TMSV becomes more squeezed. As a result, the measurement outcomes become more uniformly distributed---the variance in Eq.~\eqref{eq:r_m} scales as $\cosh 2r$ for high squeezing---and the mean in Eq.~\eqref{eq:r_a} depends less on the parameters $\phi$ and $\gamma$. In the limit of infinite squeezing, the mean does not depend on the stellar state at all, and the teleportation is perfect: Alice has the full stellar state locally. In this regime, the teleportation strategy performs just as well as DI, as is evident for large~$r$ in Fig.~\ref{f:qfi_phi_as_function_of_r}.
An optimal measurement then follows directly from DI: Alice first applies a displacement based on Bob's outcomes to complete the ideal teleportation, and then she measures $\hat  q_1 \hat q_2 + \hat p_1 \hat p_2$ (quadrature correlations) on the two modes; in the optical setting, this is equivalent to applying a 50:50 beam splitter, then
performing a photon(intensity)-difference measurement~\cite{Pearce2017optimal}. Again, this assumes she can store her state losslessly while waiting for Bob's outcomes to arrive.

As squeezing is lowered ($r \rightarrow 0$), the TMSV approaches the tensor-product state $\ket{\text{vac}}_C \otimes \ket{\text{vac}}_D$, which carries no entanglement to be exploited by Alice and Bob. (Note that this is also the limit of high loss on the TMSV; see Appendix~\ref{sec:highlosslimit}. It is discussed further in the next subsection.) In this case, Bob's measurement is equivalent to local heterodyne detection, leaving behind Alice's share of the stellar state and her share of the TMSV. The best she can do using those is local heterodyne on her share of the stellar state---the two strategies coincide, and the QFI can be calculated analytically.
In this limit, Alice's state $\op{\rho}_{\phi, \gamma}^{\vec m}$ has mean
\begin{align} 
\vec{r}_A =   \left(
\begin{array}{cc}
 -\frac{\gamma  \epsilon  \cos \phi }{\epsilon +2} & -\frac{\gamma  \epsilon  \sin \phi }{\epsilon +2} \\
 -\frac{\gamma  \epsilon  \sin \phi }{\epsilon +2} & \frac{\gamma  \epsilon  \cos \phi }{\epsilon +2} \\
\end{array}
\right)  \vec{m}, \quad \vec{r}_C = \vec{0}
,
\end{align}
and a covariance matrix $\mat \sigma_{AC}$, given in Eq.~\eqref{sigmaAChighloss}, that has no dependence on $\phi$ or $\gamma$.
 Therefore, the only contribution to the QFI for $\phi$ comes from $\vec{r}_A$, and Alice does best by using heterodyne detection to sample this mean in both quadratures;
this is because the stellar state is classically quadrature-correlated. 
Averaging over measurement outcomes, the QFI for $\phi$ and $\gamma$ is found in Appendix~\ref{sec:highlosslimit}:
\begin{subequations} \label{eq:QFIhet}
\begin{align}
\label{eq:localphi}
J^\text{tel}_{\phi} \big|_{r=0} &  = 
 \frac{2 \gamma ^2 \epsilon ^2}{\left(\left(1-\gamma ^2\right) \epsilon ^2\right)+3 \epsilon +2} =  J^\text{local het}_{\phi} ,
 \\
J^\text{tel}_{\gamma}\big|_{r=0} &=  \frac{2 \epsilon ^2}{\left(\left(1-\gamma ^2\right) \epsilon ^2\right)+3 \epsilon +2}  + \mathcal{O}(\epsilon^3) \nn
&= J^\text{local het}_{\gamma}  + \mathcal{O}(\epsilon^3).
\label{eq:localgamma}
\end{align}
\end{subequations}
The explicit form for the additional term can be found in Eq.~\eqref{app:QFIhighloss}. \color{black}
As indicated earlier, the QFI scales as $\epsilon^2$ in the weak-field limit and as $\epsilon$ for larger fields \cite{PhysRevLett.107.270402}.
This can be seen in Fig.~\ref{f:qfi_phi_as_function_of_r}, where the local heterodyne strategy sets a lower bound on the QFI for the teleportation strategy (for a given $\epsilon$).

For the phase parameter $\phi$, the lower the mean photon number $\epsilon$ in the state, the higher the required squeezing to achieve the same QFI per photon. This is qualitatively consistent with the discrete case \cite{PhysRevLett.109.070503,PhysRevLett.123.070504}, which showed that the smaller the mean photon number in the stellar state, the more distributed entanglement is required to achieve the same fidelity as the received state.
%
On the other hand, for $\gamma$, this dependence on $\epsilon$ appears less pronounced. This may be due to the fact that $\gamma$ is not encoded by a unitary and behaves differently from $\phi$.

For both $\phi$ and $\gamma$, to achieve $95\%$ of the QFI of the original stellar state, we require $r \approx 2$ (18 dB), beyond what has been achieved experimentally even for single-mode squeezing~\cite{PhysRevLett.117.110801}.

\subsubsection{QFI for \texorpdfstring{$\phi$}{phi} as a function of loss}
\label{sec:results2}

\begin{figure}
\includegraphics[trim = 0cm 0.0cm 0cm 0cm, clip, width=1.0\linewidth]{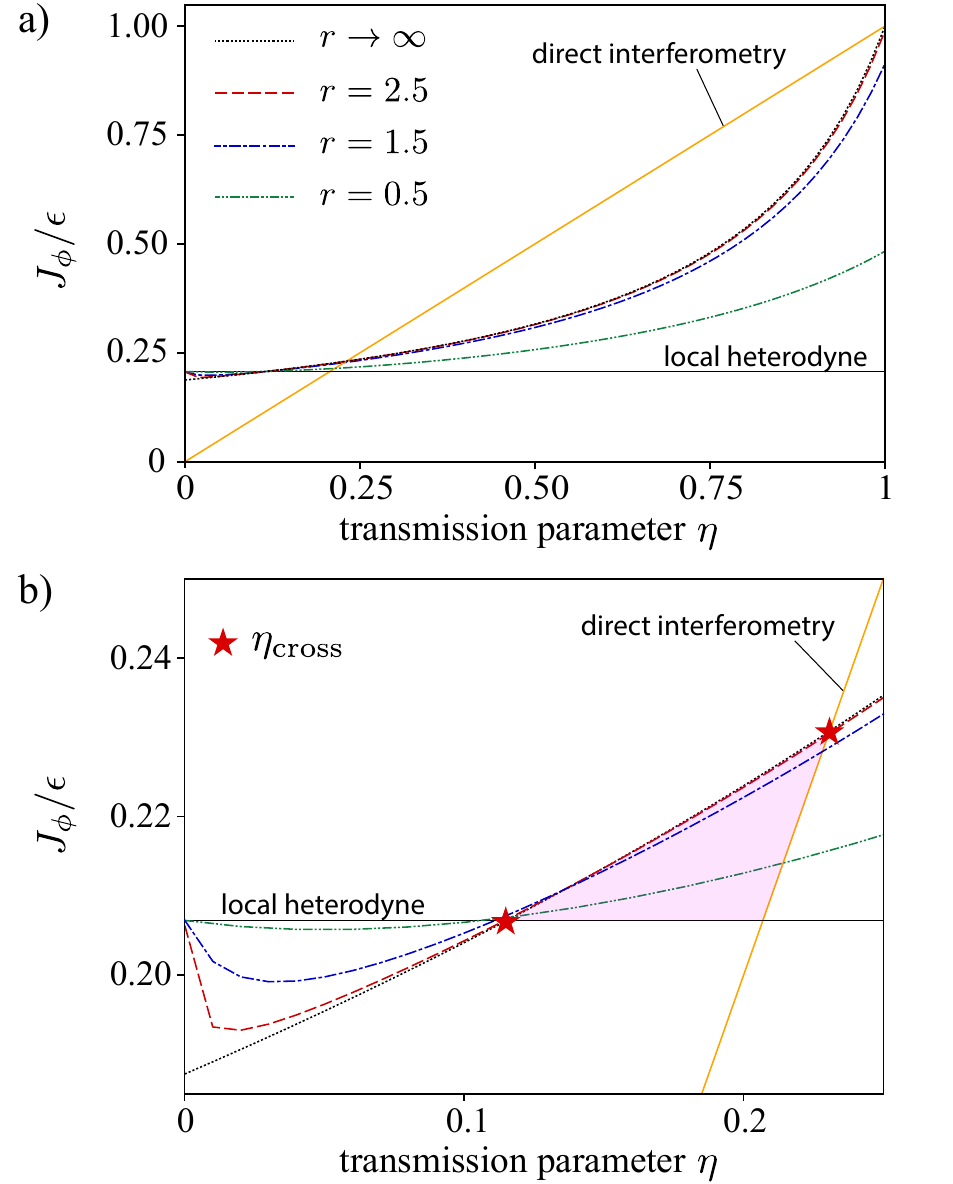} 
 \caption{\label{f:result_loss} 
 QFI per photon for $\phi$ under loss for the three strategies. Parameters are $\epsilon = 0.3$ and $\gamma = 1$. (a) QFI over all values of $\eta$. (b) A detailed view of the QFI in the high-loss regime $0<\eta<0.25$. The crossover values of the transmission parameter, $\eta_\text{cross}$, for infinite squeezing demarcate the region where the teleportation strategy outperforms the others (shaded in pink).
}
\end{figure}

We now turn to calculating the QFI per photon for the parameter~$\phi$ as a function of loss, across the three strategies. First, we show comparisons for $\epsilon = 0.3$ in Fig.~\ref{f:result_loss}. Although this value of $\epsilon$ is beyond the typical setting of weak interferometry, for which $\epsilon \ll 1$, at this value the characteristic behaviours of the three strategies are distinguishable. We return to weaker fields later. 

The benchmark is direct interferometry (DI), for which the QFI is linear in $\eta$. The local heterodyne strategy, whose classical FI is given analytically in Eq.~\eqref{eq:QFIhet}, appears as a flat line. For high values of loss, here around $\eta \sim 0.2$, a crossover occurs. At this crossover, so much light is lost transmitting the collected stellar state to a spatially local station in the DI strategy that Alice and Bob are better off simply performing heterodyne detection locally. This crossover depends on the mean photon number $\epsilon$ in the stellar state.
For the teleportation strategy, we show the QFI for several values of the squeezing parameter $r$ and for the limit of infinite squeezing $r \rightarrow \infty$, for which loss is equivalent to a random displacement channel on half of the stellar state; see Appendix~\ref{appendix:highsqueeze}.\footnote{Note that this curve is valid only for $\eta > 0$. At maximum loss, $\eta = 0$, the teleportation strategy and local heterodyne strategy coincide; see Eq.~\eqref{eq:QFIhet}.} At zero loss, the infinite-squeezing teleportation strategy performs identically to DI with other values of $r$ performing worse. 
As loss increases ($\eta < 1$), the QFI for the teleportation strategy falls faster than that for DI. This behavior is expected, because the entanglement in a TMSV is notoriously sensitive to losses. For low to moderate levels of loss, the QFI for the teleportation strategy lies between that for the DI and local heterodyne strategies.

Curiously, for higher levels of loss, a small region of ``limited quantum advantage'' exists in which the teleportation strategy can outperform both DI and the local heterodyne strategy. 
This region, highlighted in Fig.~\ref{f:result_loss}, is characterized by two crossover values of the transmission parameter that we label $\eta_\text{cross}$, one where the teleportation-strategy QFI begins to exceed DI and one where it falls below local heterodyne.
For infinite squeezing, these crossovers occur at $\eta_\text{cross} \approx 0.23$ and $\eta_\text{cross} \approx 0.11$, shown in greater detail in the inset. 
The teleportation strategy performing worse than heterodyne may be due to the fact that as the quality of the TMSV decreases, Bob's measurements serve to teleport excess noise to Alice without a compensating amount of information about $\phi$.
As $\eta$ decreases further towards zero, the QFI converges to that of local heterodyne. This is expected, because at maximal loss any TMSV becomes $\ket{\text{vac}} \otimes \ket{\text{vac}}$, the same state used to derive Eq.~\eqref{eq:QFIhet}; see Appendix~\ref{sec:highlosslimit}.

For the two extremal points, $\eta=0$ and $\eta=1$ (given $r\rightarrow \infty$), the optimal measurement operators are known \cite{Pearce2017optimal,wang2023astronomical}. For other values of $\eta$, the optimal POVMs and how to implement them physically are unknown (even assuming Alice has a perfect quantum memory to store her modes while waiting for Bob's classical outcomes). This is due to the fact that both the displacement and the correlation contain the parameters of interest -- one must measure both the displacement $\vec r_{A}$ and the correlations, proportional to $\op q_A \op q_C + \op p_A \op p_C$.

\begin{figure}
\includegraphics[trim = 0cm 0.0cm 0cm 0cm, clip, width=1\linewidth]{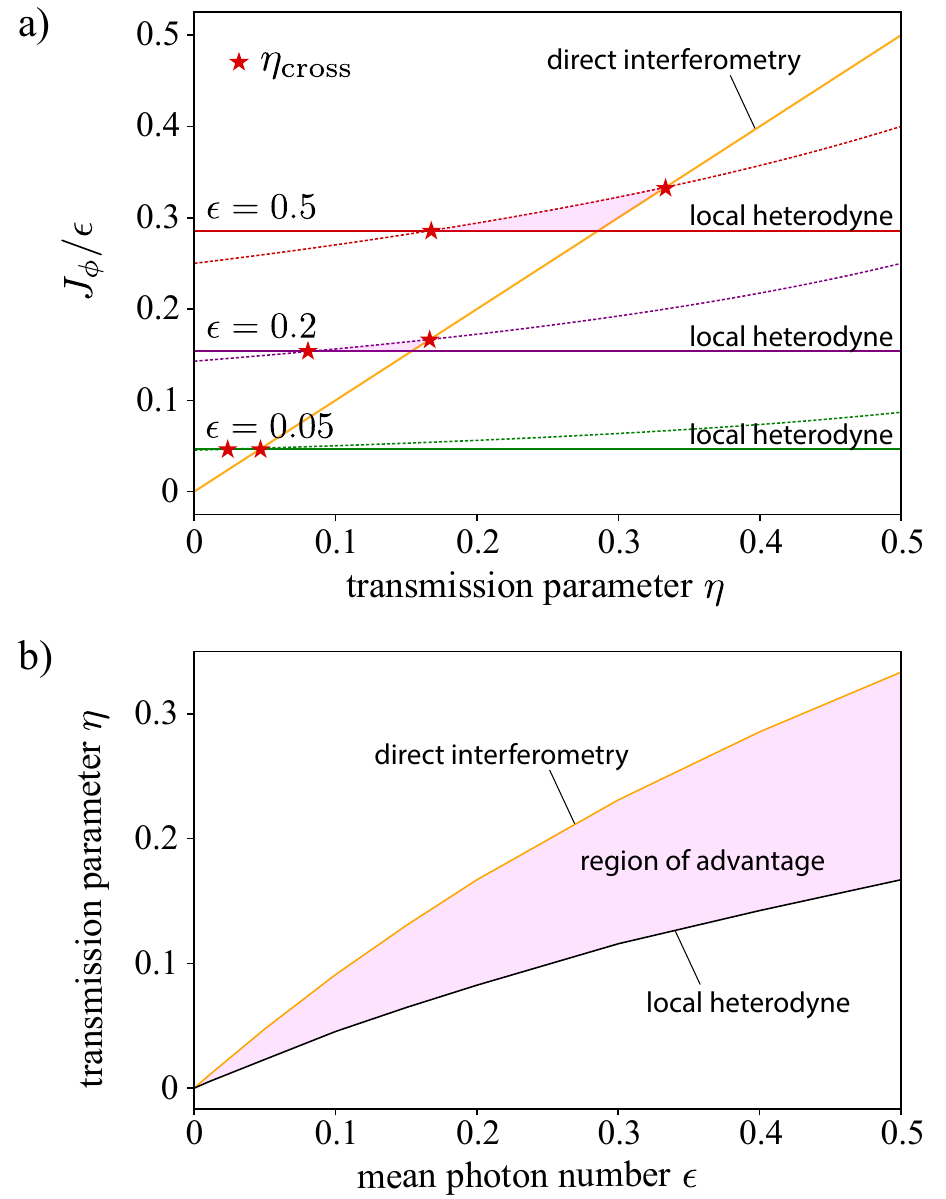} 
\caption{\label{f:inset2} 
Comparison of regions where the teleportation strategy is, in principle, advantageous for $\phi$ given different values of the mean photon number $\epsilon$. Curves for the teleportation strategy (dotted) are shown for infinite squeezing $r \rightarrow \infty$.
Red stars mark the crossover values of the transmission parameter, $\eta_\text{cross}$.
Regions of advantage for the teleportation strategy are shaded pink. 
(a) Curves for three values of $\epsilon$ showing the size of the region and the relative advantage.
(b) The region of advantage for different values of mean photon number $\epsilon$.}
\end{figure}

To quantify the region of advantage with respect to loss, we repeat the numerical analysis for other values of~$\epsilon$, including the parameter regime for weak stellar interferometry where $\epsilon \ll 1$. QFI curves for several choices of $\epsilon$ are shown in Fig.~\ref{f:inset2}(a),  using infinite squeezing for the teleportation strategy to set an upper bound on the quantum advantage. 
We see that for each value of $\epsilon$, in principle, there is reason to consider a teleportation-based estimation strategy, which can outperform both the ``classical" strategies when the expected loss is between the indicated crossover values $\eta_\text{cross}$. Figure~\ref{f:inset2}(b) extends this to show the bounding region of quantum advantage achievable with infinite squeezing.  
These figures reveal that, for weaker and weaker stellar states, the quantum advantage decreases in two ways. First, the size of the region where an advantage is possible decreases, and second, the magnitude of the advantage (in QFI) also decreases. 
Finite squeezing in any physical implementation will further reduce both the size of the region and the magnitude of the advantage. Moreover, given the increased experimental complexity associated with implementing the (unknown) optimal measurement, it is likely this small advantage will be washed out by extra noise introduced in the process.

\subsubsection{QFI for \texorpdfstring{$\gamma$}{gamma} with loss}

The QFI for $\gamma$ behaves very much like that for $\phi$ across the three strategies. We focus here on notable differences. 

\begin{figure}
\includegraphics[trim = 0cm 0.0cm 0cm 0cm, clip, width=1.0\linewidth]{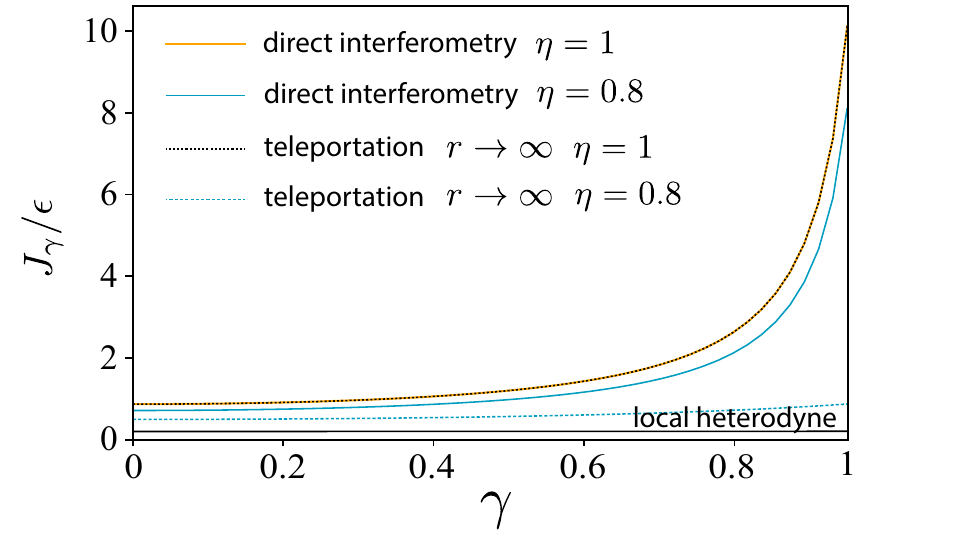} 
 \caption{\label{f:qfi_function_gamma} 
QFI per photon for $\gamma$ for parameters $\epsilon = 0.3$ as a function of $\gamma$, for $0 \leq \gamma\leq 0.95$. }
\end{figure}

The spatial parameter $\gamma$ is imprinted non-unitarily onto the stellar state and behaves qualitatively different from $\phi$: in fact, the QFI for $\gamma$ calculated directly from the stellar state is itself a function of $\gamma$. 
In the weak photon limit, its QFI is $1/(1-\gamma^2)$ per photon \cite{PhysRevLett.129.210502} and approaches infinity as $\gamma\rightarrow 1$. This is because when $\gamma=1$, the state is spatially pure: when a suitable DI measurement is made, the stellar photon will always output at the same port, and there is no error in the estimation.
The $\gamma$-dependence of the QFI for the three strategies is illustrated in Fig.~\ref{f:qfi_function_gamma}. 
The lossless, infinite squeezing teleportation strategy coincides with DI. For lower transmission, $\eta = 0.8$, a large gap between the two strategies exists, and teleportation performs only marginally better than the local heterodyne strategy. 

\begin{figure}
\includegraphics[trim = 0cm 0.0cm 0cm 0cm, clip, width=1.0\linewidth]
{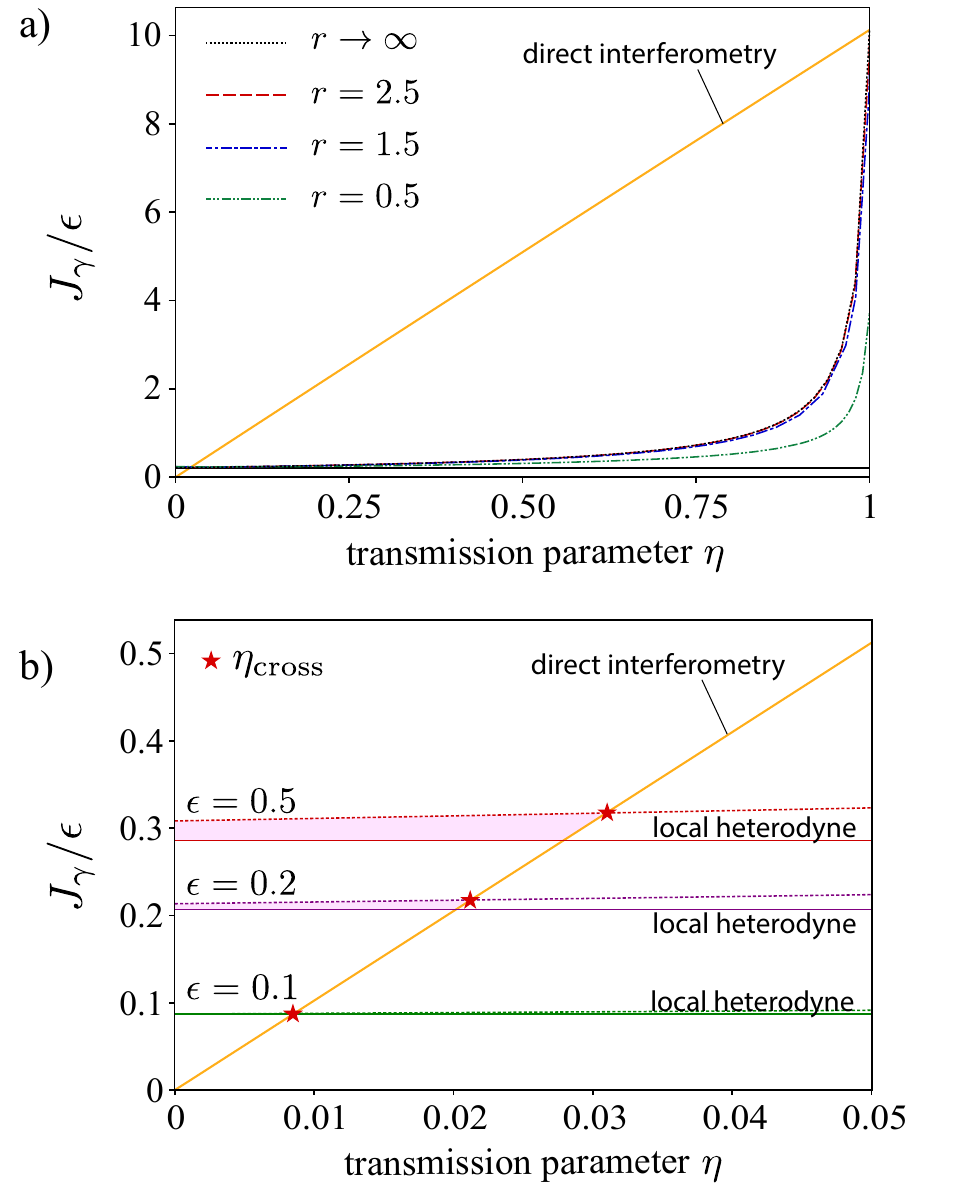} 
 \caption{\label{f:result_loss_gamma} 
Comparison of the QFI per photon for $\gamma$ under loss for the three strategies.
(a) Curves for parameters $\epsilon = 0.3$ and $\gamma = 0.95$ showing performance for several levels of squeezing in the TMSV for the teleportation strategy.
(b) Curves for several values of $\epsilon$ in the very high loss regime. The teleportation strategy performs best once the transmission parameter is below $\eta_\text{cross}$. Note that unlike the QFI for $\phi$--- compared to Fig.~\ref{f:inset2}---for $\gamma$, the teleportation strategy always performs better than the local heterodyne strategy.
}
\end{figure}

This gap between lossy DI and the lossy teleportation strategy increases as the squeezing is also lowered, as shown in Fig.~\ref{f:result_loss_gamma}(a), where we plot the QFI of $\gamma$ as a function of $\eta$ in Fig.~\ref{f:qfi_function_gamma}. 
At very high loss, the lossy teleportation scheme acquires a slight advantage compared to DI; however, the region of advantage shrinks dramatically for lower $\epsilon$; see Fig.~\ref{f:result_loss_gamma}(b). 
On the other hand, local heterodyne is consistently outperformed by the teleportation strategy, unlike the QFI for $\phi$.

\color{black}
\section{Discussions and conclusions}
\label{sec:con}

In this paper, we characterized three schemes for performing CV stellar interferometry in the full Gaussian formalism. In the lossless case, we examined the QFI of a CV teleportation-based strategy as a function of squeezing in the TMSV resource state. We found that the smaller the mean photon number $\epsilon$ in the stellar signal state, the larger the squeezing required to achieve the same QFI per photon for the parameter $\phi$. For $\epsilon \sim 10^{-2} - 10^{-3}$, we see that $\approx$ 95\% of the QFI can be achieved with $r=2$.

In the case with loss, we see that there is a transition between DI and the local heterodyne strategy: this comes from the fact that DI scales linearly in both $\epsilon$ and in the loss parameter, whereas the local heterodyne strategy is completely robust to loss (since the stellar state is measured immediately), but scales as $\epsilon^2$. The teleportation strategy requires distributed entanglement by means of the TMSV. This entanglement is very sensitive to loss, and the QFI initially drops off sharply as loss increases. However, at high loss, a small region of advantage opens up for which the teleportation strategy outperforms both lossy DI and local heterodyne strategies with respect to QFI. 

We consider this advantage to be limited for the following reasons.
First, the teleportation strategy underperforms lossy DI unless the loss is high; from Fig.~\ref{f:inset2}, for $\phi$ this occurs for $\eta \lesssim \epsilon$; for $\gamma$ this region shrinks by an order of magnitude. Even with access to very high squeezing, (a) the advantage only occurs for small regions of $\eta$ at high loss, and (b) the magnitude of the advantage is also small.
Achieving it will require measurements that are difficult to realize experimentally, and therefore we deem the quantum advantage limited.

In addition to the advantage itself being limited, we identify other obstacles to using a teleportation strategy in practice:
\begin{enumerate}
\item It requires significant two-mode squeezing.
\item It requires near-perfect quantum memories. Even then, it is likely that the optimal POVM will depend on the unknown parameters. That is because they are embedded in Alice's mean, as indicated by Eq.~\eqref{eq:r_a}. 
\end{enumerate}

For point 1, there are ways forward---notably using an entanglement distillation procedure to improve the quality of the lossy TMSV~\cite{PhysRevA.84.062309} by increasing the effective squeezing and decreasing the effective noise. The cost for this can be quite high: a quantum scissors approach~\cite{PhysRevA.100.022315} is probabilistic.
Moreover, such a distillation procedure introduces non-Gaussianity into the TMSV which must either be (a) accounted for with new analysis or (b) removed via a Gaussification process that will inject more noise.

For point 2, we further remark that for the teleportation strategy, to implement the optimal measurement, in general, quantum memories \cite{RevModPhys.83.33} will be required. This is because the post-measurement state will have non-zero displacement. Analogous to the discrete-variable case where a Pauli correction is required after teleportation, the CV states considered here will need a displacement correction that depends on Bob's quadrature measurement outcomes. Since we need to communicate this classical information from Bob to Alice, Alice will need to store her share of the quantum states losslessly. In the lossy case, the optimal measurement will depend on both the displacement and the covariance of the post-measurement state, which may make its implementation even more difficult. 

A few intriguing questions remain unanswered: First, why does the QFI for the teleportation strategy cross that of the local heterodyne strategy at a specific value of $\eta$ over many values of the squeezing parameter $r$? Second, what is the underlying mechanism that allows the teleportation strategy to outperform DI in the region of advantage after performing worse for high $\eta$? Finally, why does the teleportation strategy have a minimum QFI for each value of $r$ and then perform better as loss increases? Answers to these could unearth untapped quantum advantage with applications in sensing and quantum illumination. 

 With this work, we hope to inspire future studies that could find  strategies not only robust to noise but also experimentally achievable. A potential path forward involves correcting the loss itself during lossy DI.  
 However, it is known that Gaussian operations cannot correct Gaussian errors on Gaussian states~\cite{PhysRevLett.102.120501}. As such, to overcome loss and other Gaussian noise such as thermalization, protocols involving non-Gaussian resources may prove useful.

\medskip

\textit{Note on related work}: During the preparation of this manuscript, we learned that Ref.~\cite{wang2023astronomical} also considered CV teleportation for stellar interferometry in a complementary setting of CV repeater networks. 

\medskip

\textit{Data availability statement}: All codes used to generate the figures in this paper and a Mathematica file supporting some of the theoretical calculations are available with the arXiv posting of this paper, as arXiv ancillary files.

\begin{acknowledgments}

This paper is dedicated to Prof.~Jonathan P.~Dowling.
May the memory of him never be forgotten, and may that which he had forgotten never be remembered.

The authors thank Gavin Brennen, Alexei Gilchrist, Ryan L. Mann, Randy Lafler, R.~Nicholas Lanning, Bran Purvis, and Pieter Kok for insightful discussions. The authors acknowledge the Macquarie Centre for Quantum Engineering for funding the Jonathan P. Dowling Memorial Conference (Sydney), during which the authors initiated this research project.
ZH thanks the generous hospitality of UTS:QSI, 
where some of this work was carried out.
ZH is supported by a Sydney Quantum Academy Postdoctoral Fellowship and an ARC DECRA Fellowship (DE230100144) ``Quantum-enabled super-resolution imaging''.  
BQB and NCM acknowledge support from the Australian Research Council Centre of Excellence for Quantum Computation and Communication Technology (Project No.~CE170100012).
MMW acknowledges support from the National Science Foundation under Grant No.~2304816. 

\end{acknowledgments}

\bibliography{cluster_state}
%

\widetext
\appendix

\section{Alternate method for calculating the QFI for a Gaussian state}

Alternatively, the QFI is a function of the parameterized family $\{\hat \rho_\theta\}_{\theta}$ of states and
can be calculated using the following formula \cite[Theorem~6.3]{hayashi2006quantum}, which is useful for numerical calculations if analytical expressions are intractable:
\begin{align}
\label{eq:J}
J_\theta(\hat \rho_\theta) = \lim_{d\theta \to 0}\frac{8(1-\mathcal{F}(\hat \rho_{\theta},\hat\rho_{\theta+d\theta}))}{d\theta^2},
\end{align}
where the fidelity between two states $\hat\rho_1$ and $\hat\rho_2$~\cite{uhlmann1976transition} is
\begin{align}
\mathcal{F}(\hat \rho_1,\hat\rho_2) \coloneqq 
\operatorname{Tr} \! \left[\sqrt{ \sqrt{\hat\rho_1} \hat\rho_2 \sqrt{\hat\rho_1} }\right].
\end{align}
Thus, we can approximately evaluate the QFI numerically by choosing small $d\theta$.

When the family of states in Eq.~\eqref{eq:J} is Gaussian, the QFI can be calculated directly from the first and second moments. This follows from the fact that the fidelity between 
two Gaussian states with respective mean vectors $\vec{r}_1$ and $\vec{r}_2$ and covariance matrices $\mat{\sigma}_1$ and $\mat{\sigma}_2$ is given by~\cite{PhysRevLett.115.260501} (see also \cite{PhysRevA.61.022306,PhysRevA.86.022340}):\footnote{The formulae here differ from those in Ref.~\cite{PhysRevLett.115.260501}, because our definition for the covariance matrix differs from the one used therein.}
\begin{equation}
\mathcal{F}(\op \rho_1, \op \rho_2) = \mathcal{F}_0(\mat \sigma_1, \mat \sigma_2) \exp \!\left[ - \frac{1}{4} \vec \delta^\tp 
\left(\frac{\mat \sigma_1+ \mat \sigma_2}{2}\right)^{-1} \vec \delta \right] \label{eq:fid1},
\end{equation}
where $\vec{\delta} \coloneqq \vec r_1 -\vec r_2$, 
and
\begin{align}
  \mathcal{F}_0 (\mat \sigma_1,\mat \sigma_2)   & \coloneqq  \frac{F_\text{tot}}{\det\left(\frac{\mat \sigma_1 + \mat \sigma_2}{2}\right)^{1/4}}, 
  \\
  (F_\text{tot})^4   &\coloneqq \det \!\left[ 2\left( \sqrt{\openone + \frac{( \mat \sigma_\text{aux} \mat \Omega)^{-2}}{4}} +  \openone \right) \mat \sigma_\text{aux}\right]  , 
  \\
  \mat \sigma_\text{aux} &\coloneqq \mat \Omega^\tp\left( \frac{\mat \sigma_1 + \mat \sigma_2}{2}\right)^{-1} \left( \frac{\mat \Omega}{4} + \frac{\mat \sigma_2 \mat \Omega \mat \sigma_1}{4}  \right) .
  \label{eq:fid}
\end{align}
\color{black}

\section{Finding the teleported state}

\label{app:A}

In the following calculations, we use the notation and convention of Chapter 8 in Ref.~\cite{serafini2017quantum}.
Consider the four-mode state, before any measurements are performed, consisting of the stellar state across modes $A$ and $B$ and the TMSV across modes $B$ and $D$. The TMSV is prepared with initial squeezing $r$ and has undergone pure loss described by transmission parameter~$\eta$.

The stellar state mean vector and covariance matrix rewritten in this convention is
\begin{align} 
\label{eq:star_2}
\vec r_\star & = (
\begin{array}{cccc}
0 & 0 & 0 &0
\end{array})^\tp,
\\
\mat \sigma_\star & 
\coloneqq
\left(
\begin{array}{cc}
    \mat \sigma_a    & \mat \sigma_{ab}     \\
    \mat \sigma_{ab}^\tp & \mat \sigma_b \\
\end{array}
\right) \\
& 
=
\left(
\begin{array}{cccc}
 \epsilon +1 & 0 & \gamma  \epsilon  \cos \phi   & -\gamma  \epsilon  \sin \phi  \\
          0  & \epsilon +1                        & \gamma  \epsilon  \sin \phi  &  \gamma  \epsilon  \cos \phi \\
  \gamma  \epsilon  \cos \phi & \gamma  \epsilon  \sin \phi  & \epsilon +1 & 0  \\
 -\gamma  \epsilon  \sin \phi  & \gamma  \epsilon  \cos \phi & 0  & \epsilon +1
\end{array}
\right)
\end{align}

Using the mode ordering $(\op q_A,\op p_A,\op q_C,\op p_C,\op q_B,\op p_B,\op q_D,\op p_D )$, the covariance matrix for this state is
\begin{align} \bm \sigma_{ACBD} 
=
\left(
\begin{array}{cc}
    \mat \sigma_{\alpha}    & \mat \sigma_{\alpha \beta}     \\
    \mat \sigma_{\alpha \beta}^\tp & \mat \sigma_{\beta} \\
\end{array}
\right).
\end{align}
with block matrices 
\begin{align}
\mat \sigma_{\alpha} & \coloneqq 
\left(
\begin{array}{cc}
 \mat \sigma_a & \mat 0 \\
 \mat 0 & c \openone_2 \\
\end{array}
\right), \\
\mat \sigma_{\alpha \beta} & \coloneqq 
\left(
\begin{array}{cc}
 \mat \sigma_{ab} & \mat 0 \\
 \mat 0 & s \mat \sigma_z \\
\end{array}
\right), 
\quad
\\
\mat \sigma_{\beta} & \coloneqq 
\left(
\begin{array}{cc}
 \mat \sigma_{b} & \mat 0 \\
 \mat 0 & c \openone_2 \\
\end{array}
\right).
\end{align}
The matrices $\openone_2$ and $\mat 0$ are the two-dimensional identity and matrix of all zeros, respectively, and $\mat \sigma_z$ is the Pauli-$Z$ matrix. As in the main text, $c \coloneqq \eta \cosh(2r) + (1-\eta) $, and $s \coloneqq \eta\sinh(2r)$.

A joint EPR measurement of modes $B$ and $D$ is realized by sending them through a 50:50 beamsplitter and performing homodyne detection of the output modes---one measured in position and the other in momentum. This produces two outcomes $\vec{m} = (m_q, m_p)^\tp$, obtained with probability density~(Chapter 5 of Ref.~\cite{serafini2017quantum})
\begin{equation}
p(\vec{m}) =  \frac{\exp\left[-{\vec{m}}^\tp  [ c \openone_2 + \mat \sigma_b ]^{-1} \vec{m} \right]}
{\pi \sqrt{\det(c \openone_2 + \mat \sigma_b)}}
%
        =\frac{\exp\left(- \frac{\vec{m}^\tp \vec{m}}{(1+c+\epsilon)} \right)}{\pi (1+c+\epsilon)}.
\end{equation}
This EPR measurement can be described by a projection of the measured modes, $B$ and $D$, onto a displaced TMSV state with first and second moments,
\begin{align}
\vec{r}_m =  
\left(
\begin{array}{c}
 0  \\
 \vec{m} \\
\end{array}
\right), \qquad
\mat \sigma_m = \lim_{c' \rightarrow \infty}
\left(
\begin{array}{cc}
 c' \openone_2 & c' \mat \sigma_z \\
 c'  \mat \sigma_z & c' \openone_2 \\
\end{array}
\right),
\end{align}
a description that will be useful in calculations below. The limit $c'\rightarrow \infty$ describes infinite squeezing in the projected state---i.e., ideal homodyne detection. 

After the measurement, the remaining modes, $A$ and $C$, are projected into a conditional Gaussian state that depends on the outcome $\vec{m}$. The mean vector and covariance matrix for the post-measurement state are formally
\begin{subequations}
\label{condmoments}
\begin{align}
\vec r_{AC} &= \mat \sigma_{\alpha \beta} (\mat \sigma_{\beta} + \mat \sigma_m)^{-1} \vec{r}_m ,
\\
\mat \sigma_{AC} &= \mat \sigma_{\alpha} - \mat \sigma_{\alpha \beta} \left(\mat 
 \sigma_{\beta} + \mat \sigma_m \right)^{-1} \mat \sigma_{\alpha \beta}^\tp
 .
\end{align}
\end{subequations}
Both moments depend on the matrix inverse
\begin{align}
    (\mat \sigma_{\beta} + \mat \sigma_m)^{-1} 
    & =
    \lim_{c' \rightarrow \infty}
    \left[
\left(
\begin{array}{cc}
 \mat \sigma_{b} & \mat 0 \\
 \mat 0 & c \openone_2 \\
\end{array}
\right)
+
\left(
\begin{array}{cc}
 c' \openone_2 & c' \mat \sigma_z \\
 c' \mat \sigma_z & c' \openone_2 \\
\end{array}
\right)
\right]^{-1} \\
& \eqqcolon \left( 
\begin{array}{cc}
 \mat{M}_{11} & \mat{M}_{12} \\
 \mat{M}_{21} & \mat{M}_{22} \\
\end{array} \right),
\end{align}
which we calculate now. Block Gaussian elimination using Schur complements gives
\begin{align}
\mat{M}_{11} & = \lim_{c'\rightarrow \infty} \left( \mat \sigma_b  +  \frac{ c c' }{(c+c')}\openone_2  \right)^{-1} = \left(\mat \sigma_b + { c} \openone_2 \right)^{-1} ,
\\
\mat{M}_{12} &= \lim_{c'\rightarrow \infty} c' (\mat \sigma_b + c' \openone_2)^{-1} \mat\sigma_z \left[ (c')^2\mat\sigma_z (\mat\sigma_b + c' \openone_2)^{-1} \mat\sigma_z - (c+c')\openone_2  \right] ^{-1}
    = -\mat \sigma_z \left( \mat \sigma_z \mat\sigma_b \mat\sigma_z + c\openone_2       \right)^{-1} ,
\\
\mat{M}_{22} &= \lim_{c'\rightarrow \infty} \left[(c + c')\openone_2 - (c')^2 \mat \sigma_z  (\mat \sigma_b + c' \openone_2)^{-1} \mat \sigma_z  \right]^{-1} 
= ( \mat \sigma_z \mat \sigma_b \mat \sigma_z + c\openone_2)^{-1},
\end{align}
and $\mat{M}_{21}=\mat{M}^\tp_{12}$. Taking the limits above can be expedited using  
$(\mat \sigma_b + c'\openone_2)^{-1} \approx \frac{1}{c'}\left(\openone_2 - \frac{\mat 
 \sigma_b}{c'} \right)$.
With these relations, Eqs.~\eqref{condmoments} become
\begin{subequations} \label{condstategen}
\begin{align}
\vec r_{AC} 
&=
\left(
\begin{array}{c}
\vec r_{A} \\
\vec r_{C} 
\end{array}
\right)
=
\left(
\begin{array}{c}
- \mat \sigma_{ab} \mat \sigma_z \left( c\openone_2 +  \mat \sigma_z \mat \sigma_b \mat \sigma_z \right)^{-1} \vec{m}  \\
 s\mat \sigma_z (c\openone_2 + \mat \sigma_z \mat \sigma_b \mat \sigma_z)^{-1}  \vec{m} \\
\end{array}
\right) ,
\\
\mat \sigma_{AC} &= 
\left(
\begin{array}{cc}
 \mat \sigma_a-\mat \sigma_{ab} \left(c \openone_2+ \mat \sigma_b \right)^{-1} \mat \sigma_{ab}^\tp &
 s \mat \sigma_{ab} \mat \sigma_z\left( c\openone_2 +  \mat \sigma_z \mat \sigma_b \mat \sigma_z \right)^{-1} \mat \sigma_z  \\
 s \mat \sigma_z \left( c\openone_2 + \mat \sigma_z \mat \sigma_b \mat \sigma_z \right)^{-1} \mat \sigma_z \mat \sigma_{ab}^\tp & 
c \openone_2-s^2 \mat \sigma_z  (c\openone_2 + \mat \sigma_z \mat \sigma_b \mat \sigma_z)^{-1} \mat \sigma_z
\end{array}
 \right) .
\end{align}
\end{subequations}

Inserting the block matrices from the stellar state, Eq.~\eqref{eq:star_2}, into Eq.~\eqref{condstategen} gives the mean $\vec{r}_{AC}$ with 
\begin{align} \label{telstate:mean}
\vec{r}_A &= \left(
\begin{array}{cc}
 -\frac{\gamma  \epsilon  \cos \phi }{c+\epsilon +1} & -\frac{\gamma  \epsilon  \sin \phi }{c+\epsilon +1} \\
 -\frac{\gamma  \epsilon  \sin \phi }{c+\epsilon +1} & \frac{\gamma  \epsilon  \cos \phi }{c+\epsilon +1} \\
\end{array}
\right) \vec{m} 
,\\
 \vec{r}_C & = \left(
\begin{array}{cc}
 s (c+\epsilon +1) & 0 \\
 0 & -s (c+\epsilon +1) \\
\end{array}
\right)\vec{m},
\end{align}
and the covariance matrix
\begin{align}
\bm \sigma_{AC} = 
\left(
\begin{array}{cccc}
 1+\epsilon  -\frac{\gamma ^2 \epsilon ^2}{c+\epsilon +1} & 0 & \frac{\gamma  s \epsilon  \cos \phi }{c+\epsilon +1} & -\frac{\gamma  s \epsilon  \sin \phi }{c+\epsilon +1} \\
 0 & 1+\epsilon-\frac{\gamma ^2 \epsilon ^2}{c+\epsilon +1} & \frac{\gamma  s \epsilon  \sin \phi }{c+\epsilon +1} & \frac{\gamma  s \epsilon  \cos \phi }{c+\epsilon +1} \\
 \frac{\gamma  s \epsilon  \cos \phi }{c+\epsilon +1} & \frac{\gamma  s \epsilon  \sin \phi }{c+\epsilon +1} & c-\frac{s^2}{c+\epsilon +1} & 0 \\
 -\frac{\gamma  s \epsilon  \sin \phi }{c+\epsilon +1} & \frac{\gamma  s \epsilon  \cos \phi }{c+\epsilon +1} & 0 & c-\frac{s^2}{c+\epsilon +1} \\
\end{array}
\right),
\end{align}
for the conditional state $\op{\rho}_{\phi, \gamma}^{\vec{m}}$ held by Alice across unmeasured modes $A$ and $C$.

\subsection{High-loss or low-squeezing limit} 

\label{sec:highlosslimit}

Consider the limits $c\rightarrow 1$ and $s\rightarrow 0$ in the TMSV. This case applies to both the high loss ($\eta \rightarrow 0$) or low squeezing ($r \rightarrow 1 $) limits; see Eq.~\eqref{csparameters}. Under either of these conditions, the TMSV becomes $\ket{\text{vac}} \otimes \ket{\text{vac}}$. In the teleportation strategy, the probability density for Bob's outcomes is
\begin{align} \label{eq:highlosspr}
p(\vec m) &=    
\frac{e^{-{\vec m^\tp \vec m}/{(\epsilon +2)}}}{\pi  (\epsilon +2)}  ,
\end{align}
and Alice's post-measurement state has mean
\begin{align} \label{eq:highloss}
\vec{r}_A & = \left(
\begin{array}{cc}
 -\frac{\gamma  \epsilon  \cos (\phi )}{\epsilon +2} & -\frac{\gamma  \epsilon  \sin (\phi )}{\epsilon +2} \\
 -\frac{\gamma  \epsilon  \sin (\phi )}{\epsilon +2} & \frac{\gamma  \epsilon  \cos (\phi )}{\epsilon +2} \\
\end{array}
\right) 
\vec m,
\nn
\vec r_C &=    \vec 0    ,   
\end{align}
and covariance matrix
\begin{align} \label{sigmaAChighloss}
\mat \sigma_{AC}&= \left(
\begin{array}{cccc}
 1+\epsilon-\frac{\gamma ^2 \epsilon ^2}{\epsilon +2}  & 0 & 0 & 0 \\
 0 & 1+\epsilon-\frac{\gamma ^2 \epsilon ^2}{\epsilon +2} & 0 & 0 \\
 0 & 0 & 1 & 0 \\
 0 & 0 & 0 & 1 \\
\end{array}
\right).
\end{align}

The QFI matrix for the conditional state can be found using the expressions in Appendix~\ref{sec:qfi_matrix},
\begin{subequations}
\label{eq:fisher-heterodyne-app}
\begin{align}
J^\text{tel}_{\phi }(\op{\rho}_{\phi,\gamma}^{\vec{m}}) &= 
\frac{2 \gamma ^2 \epsilon ^2 \vec m^\tp \vec m }{(\epsilon +2) \left(\left(1-\gamma ^2\right) \epsilon ^2+3 \epsilon +2\right)}, 
\\
J^\text{tel}_{\gamma }(\op{\rho}_{\phi,\gamma}^{\vec{m}}) &=
\frac{2 \epsilon ^2 \vec m^\tp \vec m }{(\epsilon +2) \left(\left(1-\gamma ^2\right) \epsilon ^2+ 3 \epsilon +2\right)}+
\frac{4 \gamma ^2 \epsilon ^3}{\left(\gamma ^2-1\right)^2 \epsilon ^3-6 \left(\gamma ^2-1\right) \epsilon ^2-4 \left(\gamma ^2-3\right) \epsilon +8}.
\end{align}
\end{subequations}
Taking a weighted average of Eq.~\eqref{eq:fisher-heterodyne-app} by Eq.~\eqref{eq:highlosspr} gives
%
\begin{subequations} \label{app:QFIhighloss}
\begin{align}
J^\text{tel}_{\phi}(\op{\Phi}_{\phi,\gamma}) 
 & =\frac{2 \gamma ^2 \epsilon ^2}{\left(1-\gamma ^2\right) \epsilon ^2+3 \epsilon +2} ,
\\
J^\text{tel}_{\gamma }(\op{\Phi}_{\phi,\gamma}) & = \frac{2 \epsilon ^2}{\left(1-\gamma ^2\right) \epsilon ^2 +3 \epsilon +2} 
+ \frac{4 \gamma ^2 \epsilon ^3}{\left(\gamma ^2-1\right)^2 \epsilon ^3-6 \left(\gamma ^2-1\right) \epsilon ^2-4 \left(\gamma ^2-3\right) \epsilon +8}.
\end{align}
\end{subequations}
 Mathematica files supporting the calculations in \eqref{eq:fisher-heterodyne-app}--\eqref{app:QFIhighloss} are available with the arXiv posting of our paper.
 \color{black}

\subsection{Lossy, high-squeezing limit \texorpdfstring{$(\eta \neq 0)$}{(eta not equal to zero)}} 

\label{sec:lossyhighsqueeze} \label{appendix:highsqueeze}

Another extreme is for a highly squeezed state $r \rightarrow \infty$ with loss $0 < \eta \leq 1$. This excludes $\eta = 0$, where the state again returns to vacuum in both modes. 
The mean of Alice's state is
\begin{align}
\vec{r}_A &= \vec{0} , \\
\vec{r}_C & = 
 sc \mat \sigma_z \vec{m} ,
\end{align}
and the covariance matrix is
\begin{equation}  \label{app:lossyhighsqzcov}
 \mat \sigma_{AC} =  
\left(
\begin{array}{cccc}
 1+\epsilon  & 0 & \gamma \epsilon  \cos \phi & -\gamma \epsilon  \sin \phi \\
0 & 1+\epsilon & \gamma \epsilon  \sin \phi & \gamma \epsilon  \cos \phi \\
  \gamma \epsilon  \cos \phi & \gamma \epsilon  \sin \phi & 1 + \epsilon + 2(1-\eta) & 0 \\
 -\gamma \epsilon  \sin \phi &  \gamma \epsilon  \cos \phi & 0 & 1 + \epsilon + 2(1-\eta) \\
\end{array}
\right) .
\end{equation}
In the high-squeezing limit, this is equivalent to Bob's half of the stellar state being teleported to Alice with a Gaussian random displacement channel of strength $1-\eta$ applied. For no loss at all, $\eta = 1$, the stellar state is perfectly teleported to Alice.

While the mean $\vec{r}_C$ can be quite large, it has no dependence on the parameters $\phi$ and $\gamma$. The QFI is independent of the outcome $\vec{m}$ and is determined solely by the covariance matrix.

\subsection{QFI matrix for the teleported state}
\label{sec:qfi_matrix}

We find the elements of the QFI matrix for the teleported state, $\mat{J}^\text{tel}$, using Eq.~\eqref{qfi_matrix}. First, we find that $J^\text{tel}_{\phi\gamma} = J^\text{tel}_{\gamma\phi} = 0$, meaning the QFI matrix is diagonal.

Next we find the diagonal element $J^\text{tel}_{\phi \phi}$, which we simply label $J^\text{tel}_{\phi}$. The first term in Eq.~\eqref{qfi_matrix} can be calculated straightforwardly; we focus on the second term, which depends on the mean of the Gaussian state. For the teleported state here, this mean depends on the measurement outcome $\vec{m} = (m_q, m_p)^\tp$ (see Eq.~\eqref{telstate:mean}), and the second term in Eq.~\eqref{qfi_matrix} is
\begin{align}
(\partial_\phi \vec{r})^\tp \mat \sigma^{-1}_{AC}(\partial_\phi \vec{r})
 =
 \frac{2 \gamma ^2 \epsilon ^2 \left(\vec{m}^\tp \vec{m}\right) \left(c (c+\epsilon +1)-s^2\right)}{(c+\epsilon +1)^2 \left(c^2 (\epsilon +1)+c \epsilon  \left( \epsilon (1-\gamma ^2) +2\right)+c-s^2 (\epsilon +1)\right)}.
\end{align}
Taking the average over the outcomes with probability density $p(\vec m)$ according to Eq.~\eqref{eq:ensembleQFI},  with $\langle \vec{m}^\tp \vec{m} \rangle = \tr [\Cov(\vec m)] = 1+c+\epsilon$, gives
\begin{align}
(\partial_\phi \vec{r})^\tp \mat \sigma^{-1}_{AC}(\partial_\phi \vec{r})\bigr|_\text{avg}
 =
 \frac{2 \gamma ^2 \epsilon ^2 \left(c (c+\epsilon +1)-s^2\right)}{(c+\epsilon +1) \left(c^2 (\epsilon +1)+c \epsilon  \left(\epsilon (1-\gamma ^2) +2\right)+c-s^2 (\epsilon +1)\right)}.
\end{align}
Combining with the first term from Eq.~\eqref{qfi_matrix} gives us the full expression for the QFI matrix element,
\begin{align} \label{eq:phiphi}
J^\text{tel}_{\phi } 
&=
\frac{2 \gamma ^2 \epsilon ^2}{c+\epsilon +1}
\left(\frac{s^2}{c^2 (\epsilon +1)+c \epsilon  \left(\epsilon (1-\gamma ^2) +2\right)-\left(s^2+1\right) (\epsilon +1)}+\frac{c (c+\epsilon +1)-s^2}{c^2 (\epsilon +1)+c \epsilon  \left(\epsilon (1-\gamma ^2) +2\right)+c-s^2 (\epsilon +1)}\right),
\end{align}
where, again, $c$ and $s$ are defined in Eqs.~\eqref{csparameters}. 
In the limit that $s,c \rightarrow \infty$,
Eq.~\eqref{eq:phiphi} converges to  Eq.~\eqref{eq:ideal_qfi_matrix}, and in the limit that $s\rightarrow 0$ and $c \rightarrow 1$,  Eq.~\eqref{eq:phiphi} reduces to Eq.~\eqref{eq:QFIhet}.

We find $J^\text{tel}_{\gamma}$ following the same procedure,
\begin{align} \label{eq:teleported_gamma}
J^\text{tel}_{\gamma} =  \frac{X}{Y} + \frac{2 \epsilon ^2 \left(c (c+\epsilon +1)-s^2\right)}{(c+\epsilon +1) \left(c^2 (\epsilon +1)+c \epsilon  \left(\epsilon (1-\gamma ^2) +2\right)+c-s^2 (\epsilon +1)\right)} ,
\end{align}
where
\begin{align}
\label{eq:X_on_Y}
X & \coloneqq 2 \epsilon ^2 
\Big[2 \left(c^2-1\right) \gamma ^2 \epsilon  (c+\epsilon +1) \left(c^2 (\epsilon +1)+c \epsilon  \left(\gamma ^2 (-\epsilon )+\epsilon +2\right)-\epsilon -1\right)\nn
&\quad \quad -2 s^4 \left(c^2 (\epsilon +2)+\gamma ^2 \epsilon -c (\epsilon +1) \left(\left(\gamma ^2-1\right) \epsilon -2\right)\right)
+s^6 (\epsilon +2)
\nn
&\quad \quad +s^2 \big[-\left(\gamma ^4+1\right) \epsilon ^3+c^4 (\epsilon +2)-2 c^3 (\epsilon +1) \left(\left(2 \gamma ^2-1\right) \epsilon -2\right) \nn
&\quad \quad +c^2 \epsilon  \left(\left(4-8 \gamma ^2\right) \epsilon +\left(\gamma ^4-4 \gamma ^2+1\right) \epsilon ^2+4\right)+2 c (\epsilon +1) \left(\left(2 \gamma ^2-1\right) \epsilon -2\right)-4 \epsilon ^2-5 \epsilon -2\big] \Big]
,
\\
%
%
Y & \coloneqq (c+\epsilon +1) \left(\gamma ^2 \epsilon +c^2+c \left(\epsilon -\gamma ^2 \epsilon \right)-s^2-\epsilon -1\right) \left(c^2 (\epsilon +1)+c \epsilon  \left( \epsilon (1-\gamma ^2) +2\right)-\left(s^2+1\right) (\epsilon +1)\right) \times \nn
& \quad \quad \left(-\gamma ^2 \epsilon ^2+c^2 (\epsilon +2)+c \left( \epsilon ^2 \left(1-\gamma ^2\right) +4 \epsilon +4\right)-s^2 (\epsilon +2)+\epsilon ^2+3 \epsilon +2\right)
.
%
\end{align}
In the limit that $s\rightarrow 0$ and $c\rightarrow 1$, the second term in Eq.~\eqref{eq:teleported_gamma} (which is determined by the mean of the teleported state, on average) converges to the FI accessible to local heterodyne measurement (see Eq.~\eqref{eq:QFIhet}).

\color{black}

\end{document}